\newcommand{\vbb}{$0 \nu \beta \beta $}

\newcommand{\C}{$^{10}$C}
\newcommand{\Xe}{$^{136}$Xe}

\RequirePackage{fix-cm}
\documentclass[twocolumn,epjc3]{svjour3}  
\usepackage[utf8]{inputenc}
\usepackage{amsmath}
\usepackage{amsfonts}
\usepackage{amssymb}
\usepackage{calrsfs}
\usepackage[left=2cm,right=2cm,top=2cm,bottom=2cm]{geometry}
\usepackage[mathscr]{euscript}
\usepackage{xcolor}
\usepackage{graphicx}
\usepackage{subfig}
\usepackage{multicol,lipsum}
\smartqed  
\RequirePackage{graphicx}
\journalname{Eur. Phys. J. C}

\begin{document}

\title{Generative Models for Simulation of KamLAND-Zen
}


\author{Zhenghao Fu\thanksref{e1,addr1}
        \and
        Christopher Grant\thanksref{addr2}
        \and
        Dominika M. Krawiec\thanksref{addr3}
        \and
        Aobo Li\thanksref{e2,addr4,addr5}
        \and
        Lindley A. Winslow\thanksref{addr1} 
}

\thankstext{e1}{Corresponding Author. E-mail: fuzh@mit.edu}
\thankstext{e2}{Corresponding Author. E-mail: liaobo77@ucsd.edu}

\institute{Laboratory of Nuclear Science, Massachusetts Institute of Technology, 77 Massachusetts Ave, Cambridge, MA 02139, USA \label{addr1}
        \and
        Department of Physics, Boston University, 590 Commonwealth Ave, Boston, MA 02215, USA \label{addr2}
        \and
        Department of Physics, University of Warwick, Coventry CV4 7AL, United Kingdom\label{addr3}
        \and 
        Halıcıoğlu Data Science Institute, University of California San Diego, 
        9500 Gilman Dr, La Jolla, CA 92093, USA\label{addr4}
        \and 
        Department of Physics, University of California San Diego, 
        9500 Gilman Dr, La Jolla, CA 92093, USA\label{addr5}
}

\date{Received: date / Accepted: date}

\maketitle

\begin{abstract}
The next generation of searches for neutrinoless double beta decay (\vbb) are poised to answer deep questions on the nature of neutrinos and the source of the Universe's matter-antimatter asymmetry. They will be  looking for event rates of less than one event per ton of instrumented isotope per year. To claim discovery, accurate and efficient simulations of detector events that mimic \vbb is critical. Traditional Monte Carlo (MC) simulations can be supplemented by machine-learning-based generative models. In this work, we describe the performance of generative models designed for monolithic liquid scintillator detectors like KamLAND to produce highly accurate simulation data without a predefined physics model. We demonstrate its ability to recover low-level features and perform interpolation. In the future, the results of these generative models can be used to improve event classification and background rejection by providing high-quality abundant generated data. 
\end{abstract}

\section{Introduction}
\label{intro}
Event simulation is critical to modern particle and nuclear physics and is used in all experimental stages from detector design to the extraction of the final result with the corresponding statistical significance.  Traditionally, the simulation of particle detectors starts by modelling the microphysics of the particle depositing energy in the detector, and using Monte Carlo, techniques propagates the signal through the detector geometry. However,  the stochasticity and complexity of these processes makes it difficult to reproduce the detector response while simultaneously being computationally expensive to produce datasets of sufficient size. 

Rare event searches are a class of experiments that use highly specialized detectors to search for new processes that would indicate new physics at an energy-scale beyond the reach of any modern particle accelerators. Monolithic kiloton-scale liquid scintillator detectors, like KamLAND-Zen, are an excellent detector technology for rare event searches as they provide economical scaling to large volumes. For this reason, they have been the work horse of neutrino physics for many decades \cite{KamPRL,KamLAND_reactor,Borexino_CNO,Borexino_7Be1,Borexino_7Be2,snop_nd,snop_solar,dayabay,juno_summary}. KamLAND is a spherical detector, which is composed of 1\,kiloton of liquid scintillator (LS) contained in a 13-m-diameter balloon.  The LS-filled balloon is surrounded by mineral oil (acting as a buffer volume) and is viewed by 1879 photomultiplier tubes (PMTs).  A smaller secondary balloon is currently deployed at the center of the main balloon and contains LS doped with 742\,kg of $^{136}$Xe (XeLS) to search for \vbb~\cite{klz800_prl}. An observation of this rare process (current limits greater than $\sim 10^{26}$\,yrs) would prove that the neutrino is its own antiparticle, also known as a Majorana particle.  This is a key ingredient for Leptogenesis~\cite{leptogenesis}, which describes the observed matter-antimatter asymmetry in our universe.

In our previous work, we used deep learning methods to classify critical backgrounds~\cite{kamnet,nim_paper}, however the power of deep learning is not limited to background suppression. In this work, we leverage deep learning to tackle event simulation in spherical liquid scintillator detectors with the goal of producing simulations that more accurately produce the detector response while simultaneously reducing the computational burden. Ideally, we would be able to generate a large number of events from a small number of training samples realizing so-called few shot learning.

This work benchmarks two models using simulation data: the main variational auto encoder (VAE) model and an alternative generative adversarial network (GAN) model. Thanks to the data-driven nature of deep learning algorithms, the generalization to real detector data should be very straightforward. This paper is structured as follows. We first introduce the simulation of the liquid scintillator detector data used in this study in Section~\ref{section:data}. The robustness of the generative model is demonstrated by learning from these different datasets. Section~\ref{section:network_design} explains the structure of the VAE model and the GAN model. The backbone network PointNet is highlighted in this section, which elucidates the underlying mechanism of information extraction in our generative model. Finally, Section~\ref{section:training} outlines the application of the generative models and demonstrates the training results.
We evaluate the model by comparing the statistical properties of the original and generated datasets.

\section{Detector Data}
\label{section:data}
In this work, we train and evaluate our generative model on several simulation datasets. 
The first simulations are written using the Reactor Analysis Toolkit (RAT)~\cite{RAT_LOI}.  RAT is a simulation and analysis package that acts as an interface to GEANT4~\cite{geant4_1,geant4_2}. RAT is used to perform a simple KamLAND–Zen 400 simulation, referred to as sim-Fast, which is used for very fast benchmarking studies. Sim-Fast consists of events coming from \Xe~\vbb~with $Q = 2.458$\,MeV and $^{10}\text{C}$ ($\beta^+$ decay) with $Q = 3.648$\,MeV. The kinematics of the \Xe~\vbb~ events are simulated using a custom MC event generator containing momentum and angle-dependent phase space factors~\cite{kotila}. The $^{10}$C events are simulated using the default isotope decay generator in GEANT4. This correctly accounts for the long-lived first excited state of $^{10}$B, but does not include the formation of positronium.  
All events are uniformly distributed within a 3-meter-diameter mini-balloon, contained in a 13-meter-diameter balloon filled with liquid scintillator, and surrounded by a 2.5-meter-thick mineral oil buffer volume. Photons generated in each event will propagate through all these layers, reach the outer boundary of the buffer and trigger the gray-disc PMTs. The gray-disc PMT model does not produce PMT charge therefore charge is excluded from event generation associated with sim-Fast.  The photocoverage and the quantum efficiency are uniformly set to 20\% and 23\% respectively, replicating the realistic detector configuration of KamLAND-Zen 400. Secondary effects such as photon absorption, emission, and scattering are ignored in sim-Fast. The second simulation dataset is generated by KLG4sim, a detailed KamLAND simulation based on the GEANT4 toolkit.  This dataset is the standard KamLAND-Zen 800 detector MC simulation, referred to as sim-KLZ800. Compared to sim-Fast, sim-KLZ800 has been carefully tuned to replicate the response of the real detector. Thanks to the data-driven nature of deep learning algorithms, the generalization to realistic detector data should be very straightforward.

Each simulated event results in a collection of triggered PMTs. When a PMT fires, its position, arrival time (hit time t) and registered photoelectron charge (hit charge q) are recorded as a \textit{point} in the 5D space defined by the vector $[x, y, z, t, q]$. In machine learning language, a collection of points is called a \textit{point cloud}. Point cloud data has two main characteristics:
\begin{itemize}
    \item Disorder: point-cloud is insensitive to the order of points within.
    \item Invariance: point-cloud data is invariant to spatial transformations in Poincaré group.
\end{itemize}
Therefore, applying translations and rotations to the point cloud will not affect the training result. To generate the point clouds for training purpose, some additional corrections are needed. For the time dimension, two corrections are applied to calculate the proper hit time $t$ from this raw hit time. $T_{raw}$ is the raw hit time when the optical photon arrives at the PMT surface. To produce the proper hit time, the following correction is executed upon $T_{raw}$:
\begin{equation}
    t = T_{raw} - TOF - T_{0},
    \label{eqn:proper_time}
\end{equation}
where TOF is the photon time-of-flight from the event vertex to PMT position and $T_{0}$ is the proper start time of the event. By subtracting $TOF$ from $T_{raw}$, we effectively move the vertex of each event to the center of the detector. By subtracting $T_{0}$, we correct for intra-event distortion of the scintillation time profile by the vertex position. 
The calculation of $T_{0}$ is a fractional charge weighted sum of the differences between $T_{raw}$ and $TOF$ over all the PMTs. This is calculated as follows:
\begin{equation}
    T_{0}=\frac{\sum_{i}(T_{raw}^{i} - {TOF}^{i})\times q_{i}}{\sum_{i}q_{i}}
    \label{eqn:proper_start_time}
\end{equation}
where $q_i$ is the hit charge on the $i$-th PMT. We use the criteria that restrict the hit time to the event within $\pm 30\,\text{ns}$ window, and any hit times recorded outside this interval will not be considered.

Hit charge is pivotal for reproducing the energy deposition in liquid scintillator detector. It is obtained by first integrating the area under the PMT pulse, which is proportional to the number of optical photons registered at the PMT. The raw integrated value is then normalized by the the so-called 1 photoelectron (p.e.) peak integration. The 1 p.e. refers to the pulse profile where exactly one photoelectron is produced within the PMT. The calculation is displayed below:
\begin{align}
    q = \frac{\int f_{\text{event}}(\tau) d\tau}{\int f_{1\text{p.e.}}(\tau)d\tau},
\end{align}
where $f_{\text{event}}$ and $f_{1\text{p.e.}}$ are the PMT pulse and 1 p.e. pulse, respectively. This normalized value is thus the proper hit charge that reflects the number of p.e. While a low energy event creates only a single p.e., a high energy event can creates more than 1,000 p.e. PMTs with charge smaller than 0.3 p.e. are recognized as baseline fluctuations due to noise and are dropped from the point cloud. 

Up to now, we have obtained the corrected 5D point cloud dataset. However, each event (point cloud) in this dataset contains a varying number of PMT hits (points). The number of PMT hits, or the NHIT, is a essential attribute of LS detector event since it is proportional to energy. However, neural networks can only produce a fixed number of points and, therefore, cannot handle the event-wise variation of NHIT. An additional trigger-dimension is introduced to resolve this dilemma. In KamLAND-Zen, there are 1,879 PMTs spherically covering the detector surface, thus the maximal possible NHIT is 1,879. Therefore, the generative model additionally generates 1,879 floating point numbers corresponding to 1,879 PMTs in KamLAND-Zen. These floating point numbers are then fed to a sigmoid function to constrain their values between 0 and 1, and subsequently converted to 1,879 binaries using the Bernoulli distribution. If the value corresponding to a given PMT reads 1, the PMTs will be 
considered as triggered; otherwise that PMTs will be considered as untriggered.

\begin{figure}[!h]
    \centering
    \includegraphics[width=0.8\linewidth,,trim={0 0 0 0}, clip] {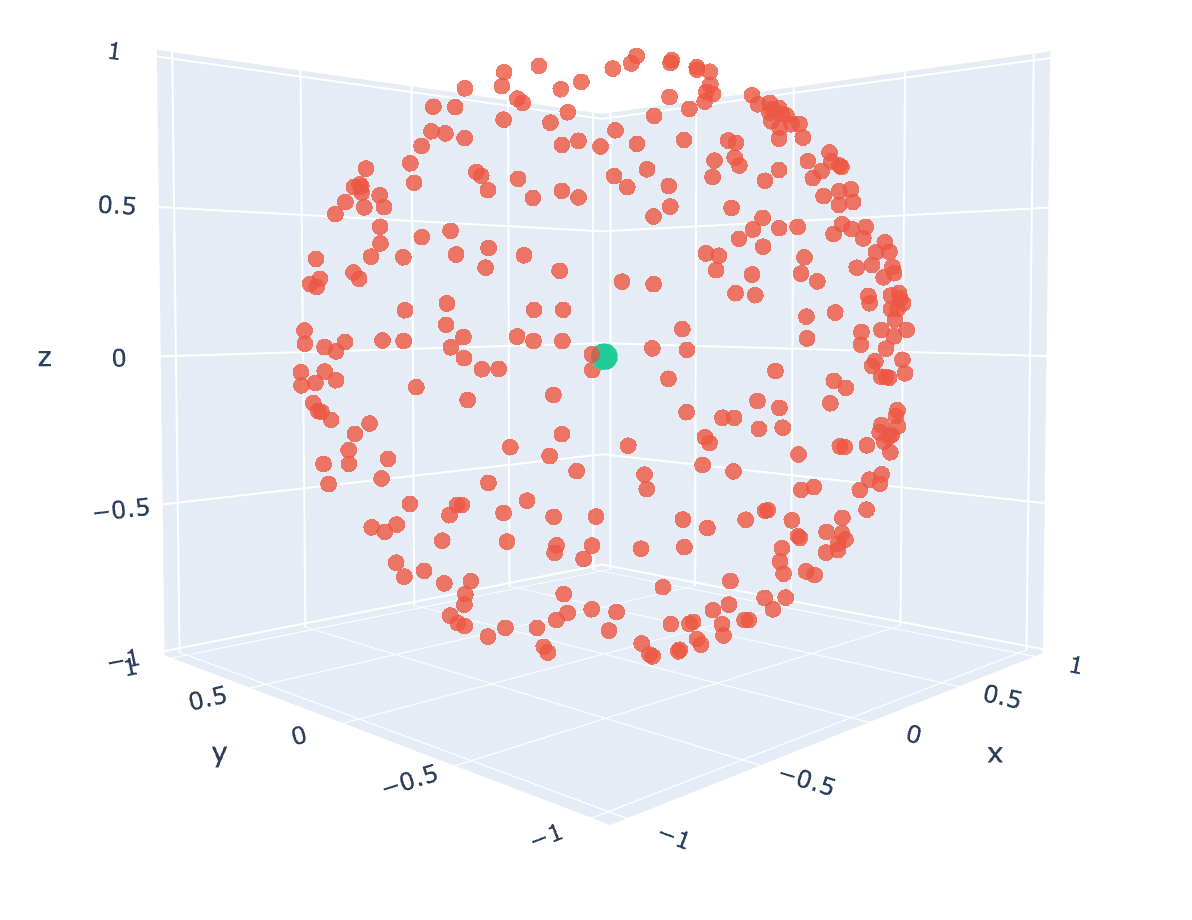}
    \caption{The inner sphere is for the untriggered PMTs (blue), and the outer sphere is for the triggered PMTs (red). The concentric double-sphere efficiently ignores the untriggered PMTs while allowing the generative model to produce a fix-sized outputs. }
    \label{figure:preprocess}
\end{figure}

The trigger-dimension is then used to transform the point clouds to a concentric double-sphere. If a PMT is marked as triggered by this dimension, its 5D values $[x,y,z,time, charge]$  are kept at the original position, while the untriggered PMTs are shifted to the origin in 5D point cloud space, as shown in Figure~\ref{figure:preprocess}. As we will discuss in Section~\ref{section:VAE}, the concentric double-sphere efficiently ignores the untriggered PMTs while allowing the generative model to produce fix-sized outputs.

Lastly, the variations of charge and time distributions are scaled to a comparable size before training. Ideally, a robust model should not be confused by scaling. It should regard time and charge values equally, however the difference in sizes of charge and time can mislead the machine. It may only focus on the dimension with a larger scale and ignore the dimension with a smaller scale. This rescaling method improves the accuracy by giving a default discrepancy in spatial information from different categories. It also speeds up the convergence of the model and prevents the gradient exploding problem during the training. 

\section{Network Design}
\label{section:network_design}
The power of a generative model emerges from its ability to probe the underlying low-level physics of the KamLAND-Zen events. If we define $X$ is the (simulated) detector events and $Y$ is the type of events. A generative model aims to describe what an observation $X$ should be when $Y$ is given. This process requires a likelihood function $P(X|Y)$ and a probability distribution $P(Y)$, where $P(Y)$ is obtained as prior knowledge, and $P(X|Y)$ is learned by the training of generative model. Two most popular generative models --- variational auto-encoder (VAE) and generative adversarial network (GAN) --- are used and compared in this study. VAE has an explicit latent space that the inference of distribution $P(X|Y)$ is enforced, while GAN has an implicit latent space and will not solve inference queries~\cite{explicit_vs_implicit}.

\begin{figure}[!h]
    \centering
    \includegraphics[width=0.9\linewidth,,trim={0 0pc 0 0pc},clip]{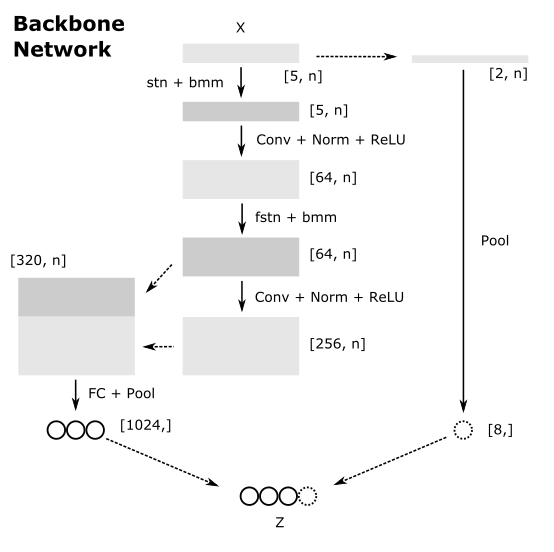}
    \caption{Major components in both generative models are shown based on the PointNet design. The $stn$ and $fstn$ are the spatial transform networks \cite{stn} used to study global information, and feature information, respectively. }
    \label{figure:pointnet}
\end{figure}

\subsection{PointNet}\label{section:backbone_network}
Typical image data such as photos or portraits can be projected on a regular pixel grid with uniform data density. However, the spatial and temporal distribution of liquid scintillator detector data is irregular and uneven and thus cannot be efficiently projected onto a 2D pixel grid. Therefore, we treat PMT hits as point clouds and adopt the PointNet model~\cite{pointnet} as the backbone network for the generative models, see Fig.\ref{figure:pointnet}. PointNet has the ability to extract both global and local features at various scales with a multi-layer structure. In KamLAND-Zen, the coverage of the PMT array is constrained by the number of PMTs. In this case, each point in the point cloud does not represent an exact photon location on the PMT photocathode, but rather marks the PMT photocathode area where a hit could occur.


\subsection{Variational Auto-encoder (VAE)}\label{section:VAE}
An autoencoder (or self-encoder) is an unsupervised learning model. A conventional autoencoder encodes data into a low-dimensional latent space representation~(or vector). This model contains a encoder network and a decoder network. The encoder network encodes the input data into a low-dimensional representation containing key information of the data. The representation is then fed into the decoder part to create an output data with the same dimensionality. During training, a reconstruction loss function is defined to minimize the differences between the output and the input data. This guarantees that important features from data can be encoded into the latent space, which can be used to reconstruct the input. Conventional autoencoders do not place any constraint upon the latent space. This model can reconstruct known events from encoded latent space vector, but lacks the ability to generate new events.

Conventional autoencoders can be upgraded to variational autoencoders~(VAE) and gain the ability to generate events. VAEs regularize the latent space representation to follow a multivariate normal distribution. This is achieved by including additional terms in the autoencoder loss function:
\begin{eqnarray*}
    L = L_0(x, \hat{x}) + \beta D_{KL}(q(z|x) \,|| \,p(z)),
\label{eqn:VAE_loss}
\end{eqnarray*}
where $L_0(x, \hat{x})$ is the reconstruction loss in the conventional auto-encoder. $D_{KL}$ denotes the Kullback-Leibler divergence~\cite{KL} computed between the returned distribution $q(z|x)$ of the latent vector $\vec{z}$ and the desired distribution $p(z)$ ~\cite{VAE_concept}. Instead of directly extracting $\vec{z}$ from the encoder network, the VAE produces two vectors with the same dimension as $z$, namely $\vec{\mu}$ and $\vec{\sigma}$. The $\vec{z}$ of the VAE is then produced using the following equation:
\begin{eqnarray*}
    z = \vec{\sigma} \cdot \vec{\omega} + \vec{\mu}
\label{eqn:reparameterization}
\end{eqnarray*}
where $\vec{\omega}$ is sampled from a multivariate normal distribution with the same dimension as $\mu$ and $\sigma$. This method is referred to as a ``reparameterization trick''. It allows gradient to flow through the network. After training, we can repetitively sample from the multivariate normal distribution to simulate new events. Lastly, $\beta$ is the hyperparameter that controls the strength of the latent space regularization. 
\begin{figure}[!h]
    \centering
    \includegraphics[width=0.9\linewidth,,trim={0 0pc 0 0pc},clip]{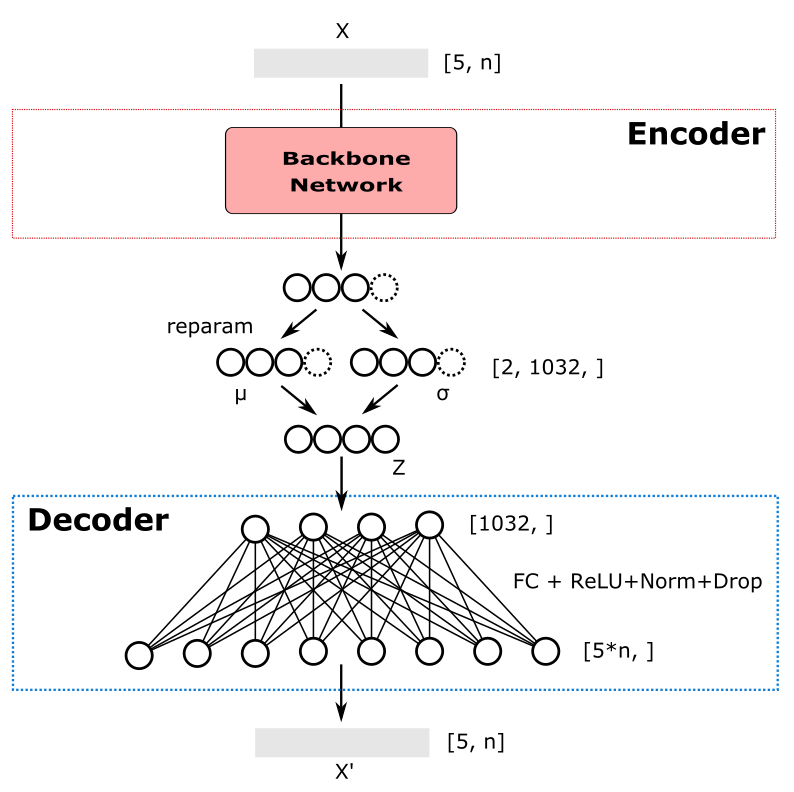}
    \caption{Schematic diagram of our generative model based on the VAE architecture. The encoder transforms the original input $X$ to a latent vector $Z$, and the decoder recovers the input $X'$ from this latent vector.}
    \label{figure:VAE}
\end{figure}

The structure of the customized VAE for KamLAND-Zen data is displayed in Figure~\ref{figure:VAE}. The encoding part is the PointNet model we introduced in previous section, and the decoding part is a fully-connected neural network. In this work, the reconstruction loss $L_0$ contains two parts. The first part is the Chamfer distance~\cite{chamfer_distance} calculated between the input and output data. Chamfer distance is defined as the sum of the minimal distance between each pair of points separately from two point clouds. As discussed in Section~\ref{section:data}, the untriggered PMTs will be at the origin after the concentric double-sphere transformation. When calculating the Chamfer distance, contributions from the untriggered PMTs will be 0 thus does not contribute to the network training. Therefore, we are effectively training on triggered PMTs when minimizing the Chamfer distance. Furthermore, we use the BCE loss to regularize the trigger-dimension. This loss effectively limits the total number of triggered PMTs in the output data, compensating for the Chamfer loss's neglect of the inconsistent number of hits in the output and input.

\subsection{Generative Adversarial Network (GAN)}\label{section:GAN}
The GAN takes a different approach to generate events. The GAN consists of two networks competing against each other: the discriminator $D(x)$ is designed to determine the authenticity of data $x$, and the generator $G$ converts the randomly-sampled noise vector $z$ to a synthetic event $G(z)$. In a well-trained GAN model, the discriminator and generator will reach a Nash equilibrium thereby leading to efficient event generation from random noise $z$.

The Wasserstein-GAN model \cite{WGAN} is adopted to avoid the divergence in losses of generators and discriminators. Wasserstein-distance measures the earth-moving distance between the distributions of real and fake events, which provides a valid gradient for the GAN model to train. A gradient penalty \cite{gp} is also implemented to facilitate the training, which gives a bound on the Lipschitz-norm of the gradient of the discriminator function $D(x)$ and limits the discriminator from making dramatic changes when the input sample only varies slightly. 

\begin{figure}[!h]
    \centering
    \includegraphics[width=1.0\linewidth,,trim={0 0pc 0 0pc},clip]{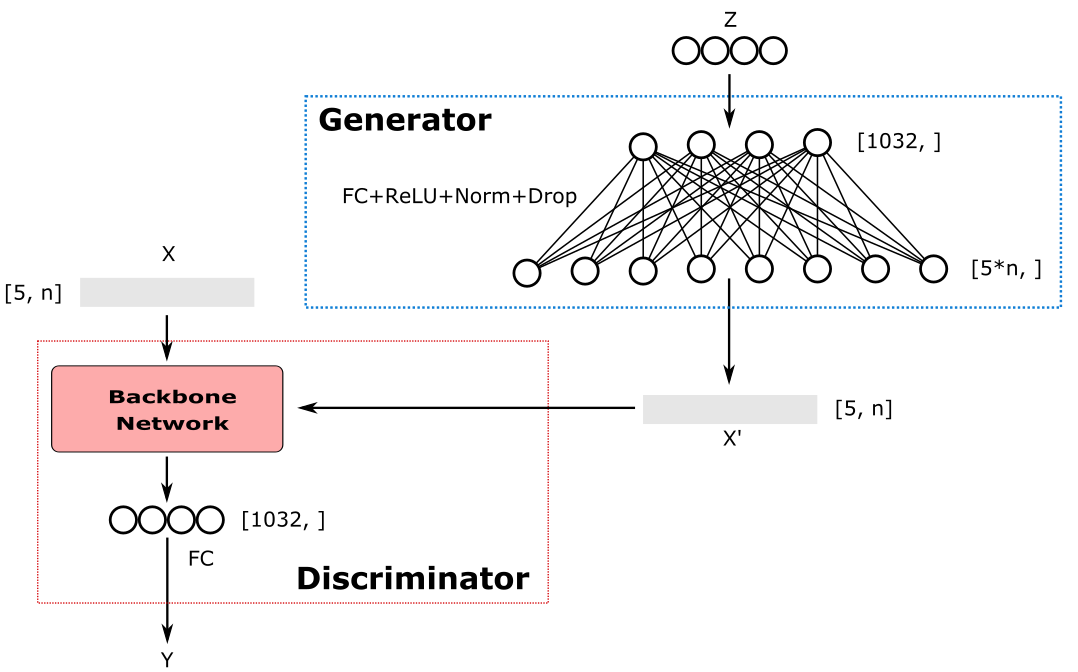}
    \caption{Schematic diagram of our generative model based on the GAN architecture. Original input $X$ and generated input $X'$ from random noise $Z$ are compared and classified. }
    \label{figure:GAN}
\end{figure}

In this work, we developed the GAN to fulfill the same event generation task of the VAE. However, the reconstruction power of GAN turns out to be worse than the VAE model in~\ref{section:VAE}, as shown in Table~\ref{Tab:WD}. The design of  GAN is shown in the Fig.~\ref{figure:GAN}.

\section{Training Results}
\label{section:training}
We used several datasets to train and validate the generative models. The performance of conventional autoencoder, VAE and GAN models are shown for both sim-Fast and sim-Full datasets. Lastly, we also demonstrate VAE's capability to conduct few-shot learning with merely 50 training samples.

\subsection{Autoencoder Result}
\begin{figure}[!h]
    \centering
    \includegraphics[width=1.0\linewidth,,trim={0 0pc 0 0pc},clip]{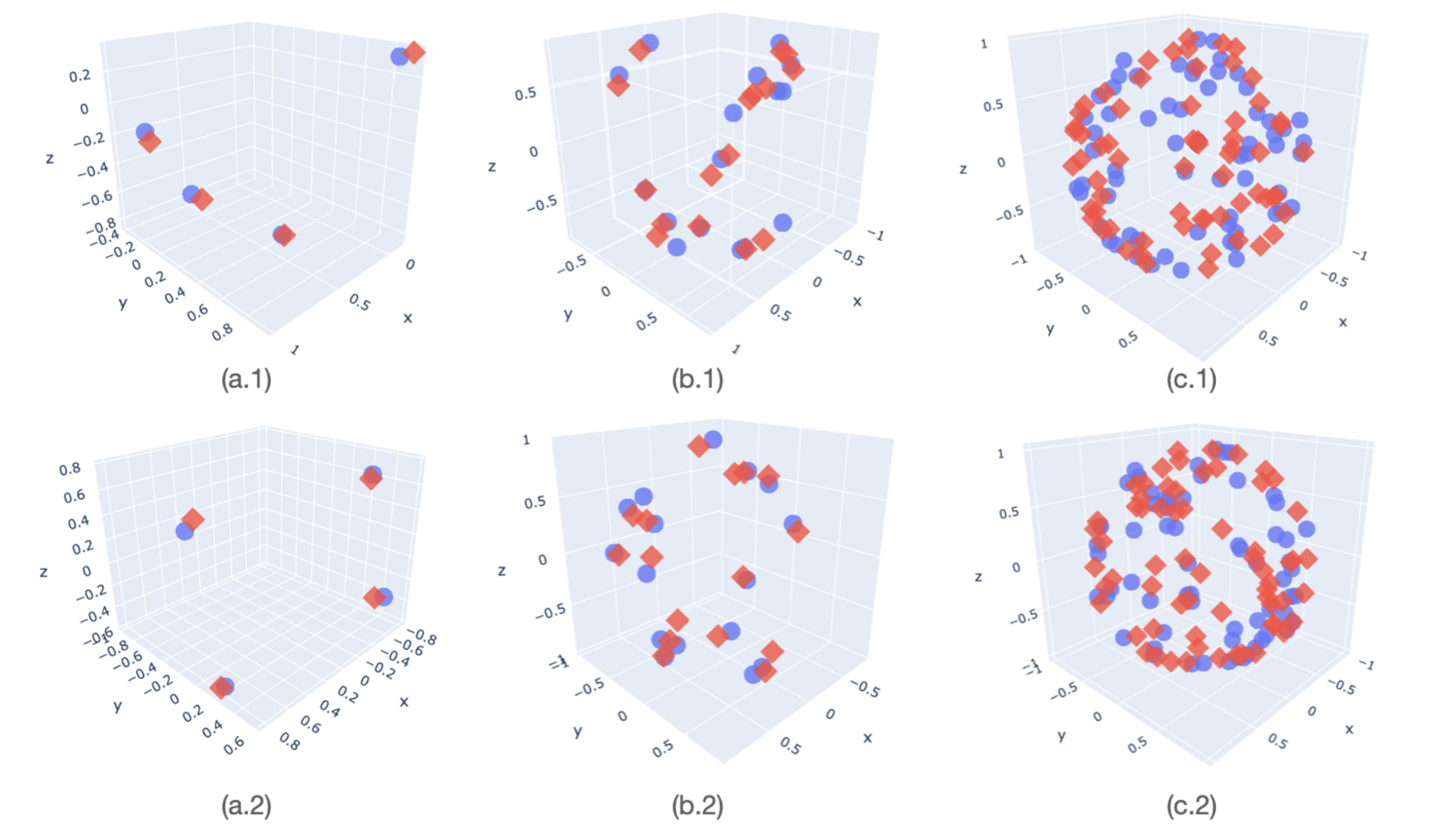}
    \caption{Hit point location reconstruction with NHIT = 4, 16, and 64. The lower row plots are training samples, and the upper plots are testing samples. The blue circles are real data, while the red diamonds are the reconstructed.  }
    \label{fig_auto_loc}
\end{figure}

The results from the conventional autoencoder are visualized in Figure~\ref{fig_auto_loc}. Instead of using a concentric double-sphere, we used a fixed size of point cloud to test the performance of conventional autoencoder. A given number of triggered PMTs are randomly selected to form the input point cloud, and the conventional autoencoder model is trained to generate the same number of points at the decoder output. With the increased number of points, the conventional autoencoder can effectively record and reconstruct the spatiotemporal information of all PMTs. As seen by the similarity between the data points and the generated points in Figure~\ref{fig_auto_loc}, the autoencoder does a decent job reconstructing NHITs.

\subsection{Benchmarking Dataset}
\label{subsec:benchmark}
\begin{figure*}[!h]
    \centering
    \includegraphics[width=0.85\linewidth,,trim={0 0pc 0 0pc},clip]{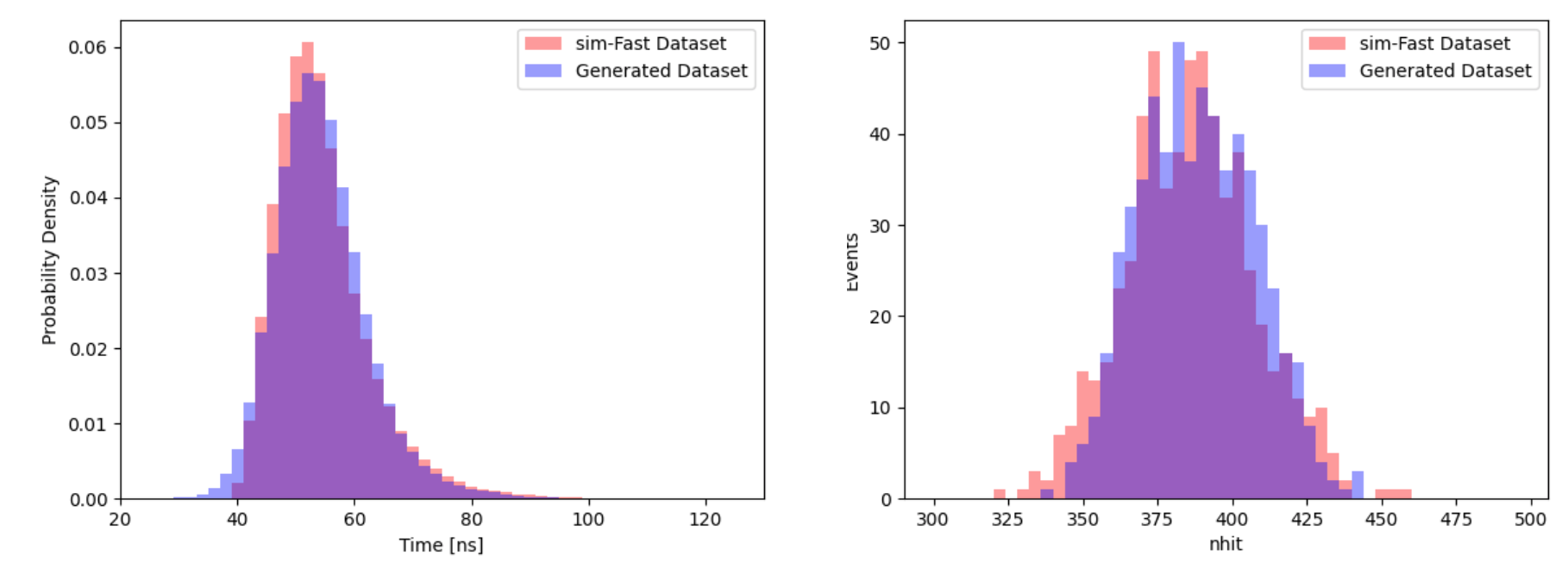}
    \caption{Left: time distribution of all triggered hit points in each event; right: nhit distribution for all \Xe~events. }
    \label{fig:benchmark_time}
\end{figure*}

\begin{figure*}[!h]
    \centering
    \includegraphics[width=0.65\linewidth,,trim={0 0pc 0 0pc},clip]{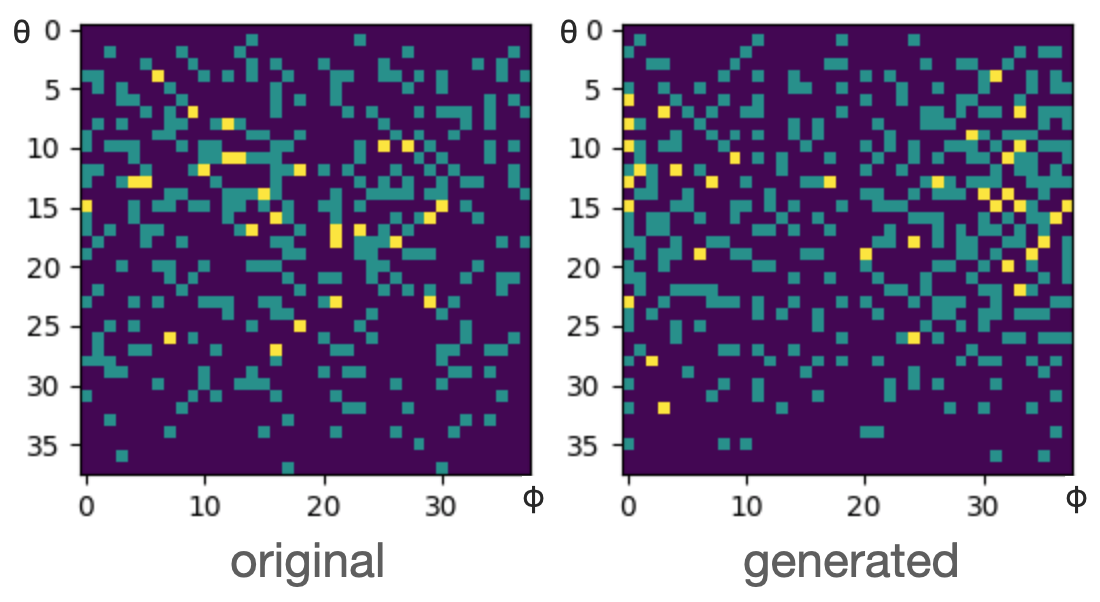}
    \caption{Hit maps of \Xe~\vbb~events in the $\theta$ and $\phi$ dimensions for the real data (left) and generated data (right). }
    \label{fig:benchmark_hit}
\end{figure*}

Next we use sim-Fast to demonstrate the power of VAE. As discussed in Section~\ref{section:data}, sim-Fast does not simulate PMT charge, therefore the event generation only outputs PMT time and the trigger-dimension. As seen on the left side of Figure~\ref{fig:benchmark_time}, the hit time distribution of generated \Xe~and~\C~events is identical to the input sim-Fast events. The number of triggered PMTs~is also accurately generated as shown on the right side of Figure~\ref{fig:benchmark_time}. The comparison between the real hit map and generated hit map is shown in Figure~\ref{fig:benchmark_hit}. To make the hit map, all triggered PMTs are projected onto a $38\times38$ grid by opening up the sphere along the $\theta$ and $\phi$ dimension. The generated hit map shows sharp PMT hit patterns without smearing effects or unrealistic hits.

\subsection{Standard Simulation of \vbb~Decay}
\label{subsec:standard}
\begin{table*}[!h]
\centering
  \begin{tabular}{|c|c c|cc|}
    \hline
    $J/\mathcal{R}$ & VAE IoU[\%] & VAE WD[a.u.] & GAN IoU[\%] & GAN WD[a.u.]  \\
    \hline
    Time Dist. & 89.0 & $9.32\times 10^{-2}$ & 85.5&0.241 \\
    Charge Dist. & 91.9&0.254 & 80.0&0.461 \\
    Nhit Dist. & 81.2&1.96 & 76.2&2.98 \\
    Tot Charge Dist. & 76.0&5.57 & - &-\\
    \hline
  \end{tabular}
  \caption{Comparison table between the VAE and GAN models. The comparison is performed with two metrics: Intersection over Union~(IoU) and normalized Wassertein distance~(WD). The GAN is trained with 10,000 epochs. }
  \label{Tab:WD}
\end{table*}
To demonstrate the power of generative models on sim-Full, we train VAE and GAN to simultaneously generate the trigger-dimension, time and charge for every PMT. The generation performance is evaluated with Intersection over Union (IoU) metric, also named as \textit{Jaccard index}, between the original ($P_0$) and the generative distributions ($P_g$). The IoU ratio is calculated using the binned distributions of the input and output data as:
\begin{equation}
    J(P_{0}, P_{g}) = \frac{\sum_{k=1}^n min(P_0\left[k\right], P_{g}[k])}{\sum_{k=1}^n  max(P_0[k], P_{g}[k])},
\end{equation}
where $k$ is the index of histogram bins of both the original and generated data. The ratio $J$ ranges from 0 to 1. Two identical datasets will have $J=1$, and any difference in distribution will result in a decrease in IoU. Meanwhile, we also included normalized Wasserstein distance to evaluate the generation accuracy, expressed as: 
\begin{equation}
    \mathcal{R}(P_{0}, P_{g}) = \frac{W(P_{0}, P_{g})}{\max(D_0)-\min(D_0)} ,
\end{equation}
where $D_0$ is the input data. Two identical datasets will also have $\mathcal{R}=0$, and any difference in distributions will increase this value.

\begin{figure*}[!h]
    \centering
    \includegraphics[width=0.85\linewidth,,trim={0 0pc 0 0pc},clip]{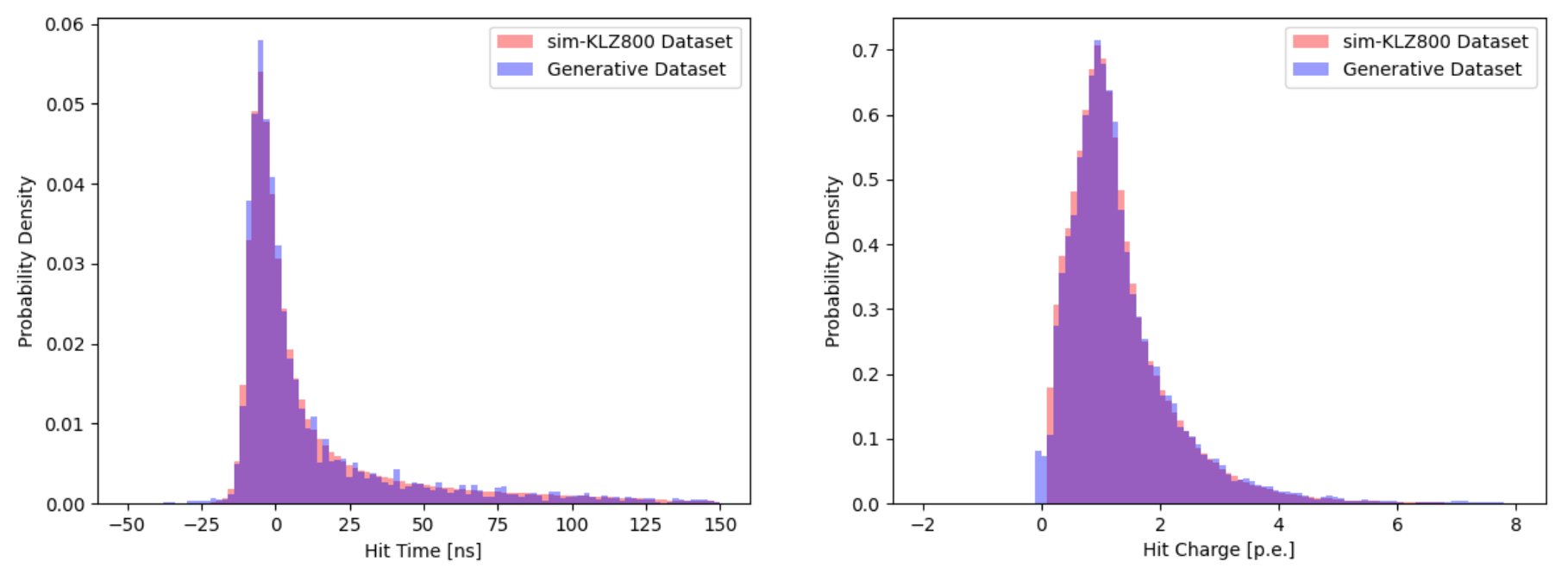}
    \caption{The averaging time distribution of all triggered hit points is shown on the left, and the averaging charge distribution of all triggered hit points is shown on the right. }
    \label{fig:std1}
\end{figure*}

\begin{figure*}[!h]
    \centering
    \includegraphics[width=0.85\linewidth,,trim={0 0pc 0 0pc},clip]{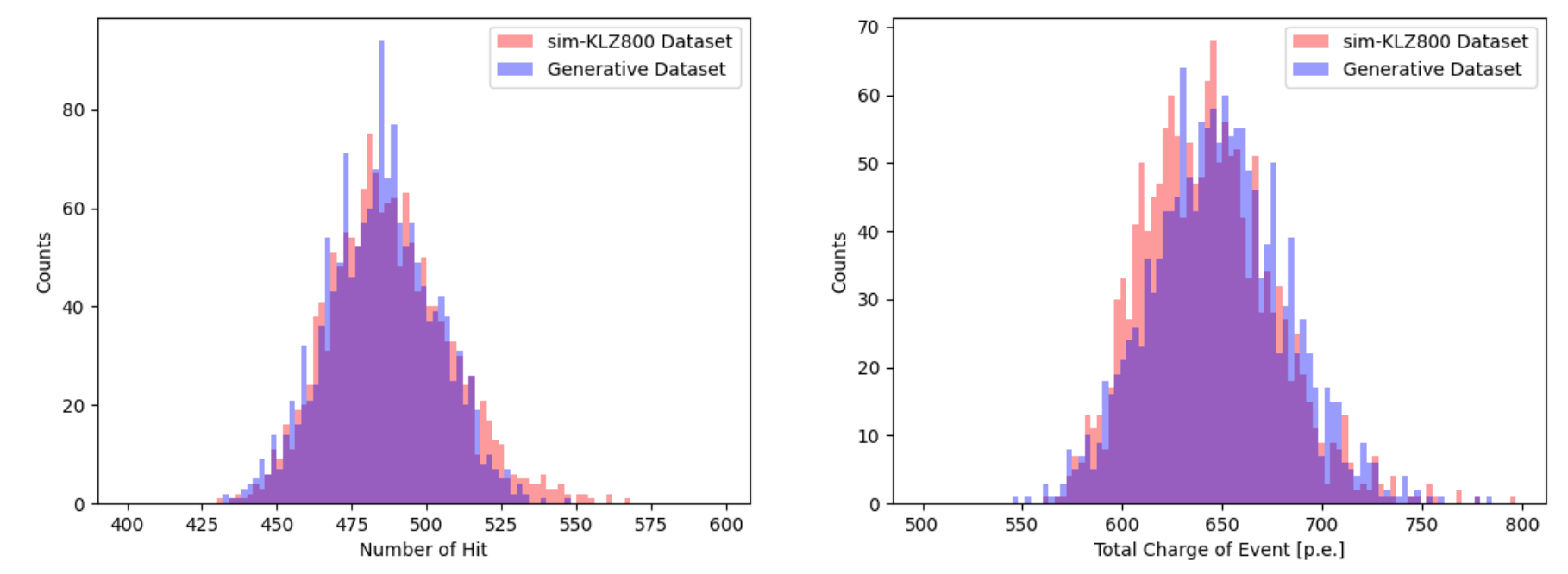}
    \caption{The NHIT distribution for 1000 \Xe~\vbb~events is shown on the left and the total charge distribution for the same events is shown on the right. }
    \label{fig:std2}
\end{figure*}

The training result is illustrated in Figure~\ref{fig:std1}, where the distributions of hit times and hit charges of 1000 events are compared. 
The hit time has $\color{black}J_t=89.01\%$, and the charge has $\color{black}J_q=91.95\%$. This indicates a good agreement between the real and generative events, and the difference can be considered as the statistical fluctuations. The NHIT and total charge of original events and generated events are compared in Figure~\ref{fig:std2}. The NHIT distribution has $\color{black}J_{N}=81.21\%$, and the total charge distribution has $\color{black}J_{Q}=75.95\%$. We also tested the GAN models on the same dataset, and the training result is shown in Table \ref{Tab:WD}. The GAN model gives decent performance, but it underperforms when compared to the VAE on the four distributions under both evaluation metrics.

\subsection{Few-Shot Learning}
One essential advantage of generative model is its few-shot learning capability, in that it learn crucial features with an extremely small number of training events, denoted as \textit{few-shot dataset}. To perform few-shot learning, we first pre-train our generative model using a large $^{214}$Bi dataset from sim-KLZ800. Pre-training allows the generative model to learn basic features of liquid scintillator events and provide a good starting point for training. We selected $^{214}$Bi for pre-training because a pure set of $^{214}$Bi can be easily selected by delayed coincidence tagging in KamLAND-Zen. Next, the pre-trained model is trained with a relatively small collection of 50 \Xe~\vbb~events forming the few-shot dataset. The trained model is used to generate 1000 \Xe~\vbb~events and compared to 1000 real \Xe~\vbb~events.

\begin{figure*}[!h]
    \centering
    \includegraphics[width=0.85\linewidth,trim={0 0 0 0},clip]{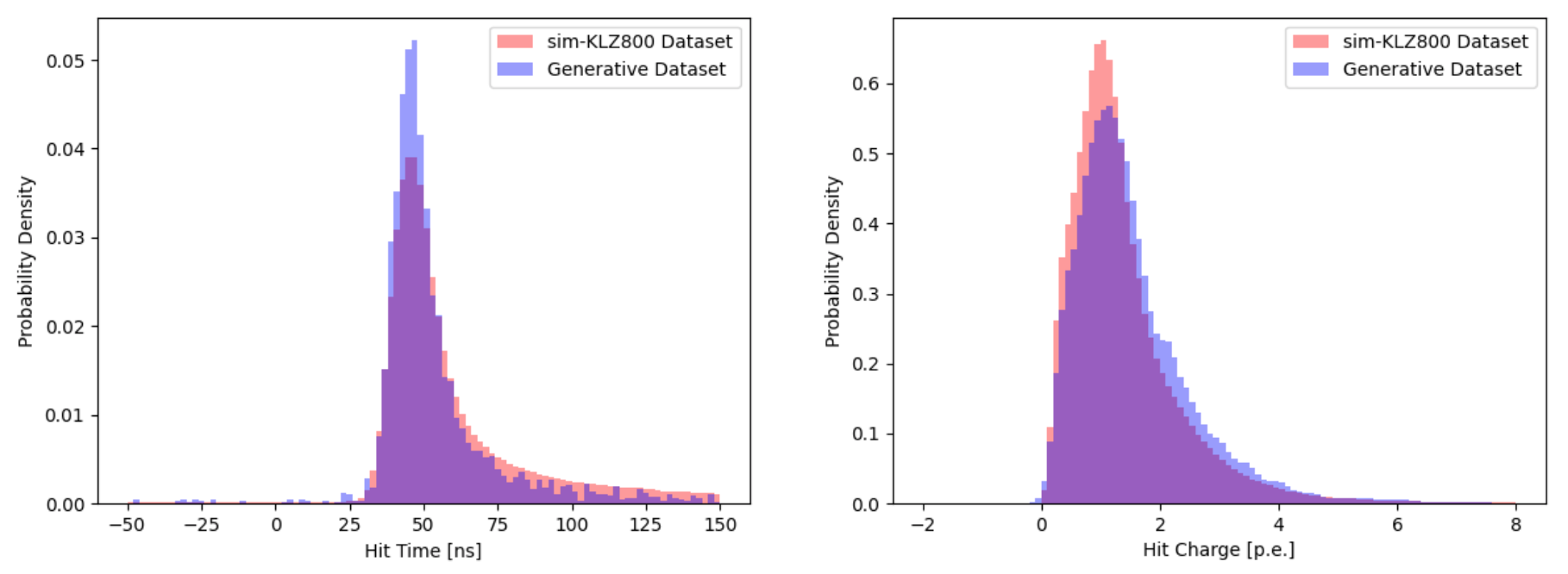}
    \caption{Few-shot training result forf the reconstructions of time and charge distributions.}
    \label{fig:fewshot}
\end{figure*}

The result of the few-shot training is shown in Figure~\ref{fig:fewshot}. With an extremely small number of few-shot samples, few-shot learning reproduces the statistical distribution of individual events where the associated uncertainties for IoU values are obtained through bootstrapping. The IoU values for time distribution $\color{black}J_t=92.710(38)\%$ and charge distribution $\color{black}J_q=92.410(44)\%$ are approximately equal to normal learning, while the network without few-shot training gives $J_t=88.918(53)\%$ and $J_q=90.589(56)\%$. In addition, few-shot learning also reproduces accumulated statistics.
The IoUs are $\color{black}J_{N}=79.067(826)\%$ and $\color{black}J_{Q}=77.581(761)\%$ for NHIT and total charge distributions, respectively, which is a significant improvement over the network without few-shot training that only gives $J_{N}=29.892(176)\%$ and $J_{Q}=33.283(122)\%$. This result dispels the concerns that the input dataset is too small to have a clean distribution over the cumulative attributes.

In KamLAND-Zen 800, we found an increase in the background rate at the inner balloon bottom, possibly due to the settling of dust particles. However, an unambiguous identification of the source is impossible due to limited statistics~\cite{klz800_prl}. In this case, the few-shot data are the unknown background events with limited statistics. Leveraging few-shot learning, we will be able to boost the population of few-shot data using the following steps: we first use tagged $^{214}$Bi data to pre-train the VAE, then use the collected few-shot data to train the network. After training, the generative model will be able to generate as many events as needed and they can be used to understand the origin of a small number of spurious events in real data. Note that the training procedure is based on detector data, therefore the generated events will accurately include effects coming from the real detector response. 

\section{Conclusion and Outlook}
\label{sec:conclusion}
Traditional simulation is based on inferences from first principles, but it depends on the accuracy of the input parameters and often fails to exactly reproduce the detector microphysics. In this work, we developed two generative models to simulate liquid scintillator detector data. The generative models improve the efficiency in generating data with a decent reconstruction accuracy. With a standard detector configuration similar to the current KamLAND–Zen detector, the variational autoencoder model can accurately simulate data with $\color{black} J \gtrsim 90\%$ and $\mathcal{R}\lesssim 5\%$. Furthermore, we examine the possibility of few-shot learning using the given generative model. With fewer than 50 training events and an easy-to-collect pre-training dataset, our generative model can extract critical features from the training events to significantly boost its population.  

This work's focus is the optimization of the algorithms to study the statistical properties of raw data and generate like-real detector events. However, it is well-known that the VAE-based generative model generally falls short on exploring the full latent space. This limitation sacrificed the diversity of data to ensure the validity of the generative data. In future studies, we plan to leverage new deep learning models, such as the diffusion model~\cite{diffusion_model}, to improve on the VAE and generate events with a broader distribution. These studies will benefit particle and nuclear physics by offering a faster and data-driven method for simulation development.

\section{Acknowledgments}
This material is based upon work supported by the National Science Foundation under Grant Numbers 2110720, 2012964. This work is done in support of the KamLAND–Zen experiment and we thank our collaborators for their input. We thank Alexander Leder and Daniel Mayer for their review of the manuscript. The KamLAND-Zen experiment is supported by JSPS KAKENHI Grant Numbers 21000001, 26104002, and 19H05803; the Dutch Research Council (NWO); and under the U.S. Department of Energy (DOE) Grant No. DE-AC02-05CH11231, as well as other DOE and NSF grants to individual institutions. This research was performed, in part, using the Boston University Shared Computing Cluster.

\newpage

\bibliographystyle{spphys}
\bibliography{VAE}

\end{document}